\documentstyle[aps,twocolumn,epsf]{revtex}
\begin{document}

\draft

\title{
Purifying two--bit quantum gates and joint measurements in 
cavity QED
}

\author{S.J.~van Enk, J.I.~Cirac, P.~Zoller}

\address{
Institut f\"ur Theoretische Physik, Universit\"at Innsbruck,
Technikerstrasse 25, A--6020 Innsbruck, Austria.
}

\date{\today}

\maketitle

\begin{abstract}
Using a cavity QED setup we show how to implement a particular
joint measurement on two atoms in a fault tolerant way. Based on
this scheme, we illustrate how to realize quantum communication
over a noisy channel when local operations are subject to errors.
We also present a scheme to perform and purify a universal two--bit
gate.
\end{abstract}

\pacs{PACS: 03.65.Bz, 42.50-P}

One of the most intriguing features of quantum mechanics is the
possibility of entangling physical systems, which has both practical and
fundamental implications. On the one hand, Bell's theorem \cite{Bell}
states that quantum mechanics and any local realist theory are
incompatible based on the peculiar properties of entanglement. On the
other, quantum communication and computation exploit these properties to
guarantee secure communication and to construct algorithms that allow
fast computations \cite{applications}. 

In a series of remarkable experiments the first steps towards these
lofty goals have been taken \cite{Wi95,Kimble,Ha97}. In particular, it
seems that quantum communication will have several practical
applications in the near future. For example, quantum cryptography has
been tested experimentally over long distances using standard
telecommunication fibers \cite{QCryptoexp}. This, combined with recent
proposals \cite{QN,En97} for exchanging quantum information between
atoms and photons based on cavity QED suggests that a full quantum
network including local processing and transmission of quantum data is
possible. Since practical uses of quantum networks require a high degree
of entanglement one might think that this is not feasible due to the
presence of errors and decoherence. However, the recent discovery of
quantum error correction protocols and purification schemes
\cite{qerrorc,Scien,Purification} shows that this is not a fundamental
obstacle. In a quantum network one can classify the errors in two
categories: {\em transmission errors}, i.e. those occurring during
transfer of quantum information between nodes and {\em local errors},
i.e. those occurring during local processing and measurements. Since
transmission errors are much more likely than local errors, one usually
assumes that the latter are absent. With this assumption, noisy channels
have been defined and protocols have been devised to achieve ideal
transmission of quantum information \cite{Schumacher}. Most proposals
allow for quite general types of noise, and require infinitely many
resources to achieve this goal. In contrast, based on a specific model
for quantum communication we have proposed a protocol that requires only
finite resources \cite{En97} and corrects for the physically relevant
errors. Therefore, in that physical scenario the only remaining problem
is local errors. Although one could in principle use standard error
correction schemes to solve this problem, this would again require
infinite resources.

In this work we will give a physical implementation that allows to
perform local operations and measurements ideally using finite
resources. The scheme is based on cavity QED and therefore can be easily
connected to the previous proposal for quantum communication
\cite{QN,En97}. We will assume that operations acting on a single atom
are error free, whereas any other operation is not. This is motivated by
the experimental fact that single bit operations are much simpler than
multiple bit operations \cite{Ha97}. First we will show how to perform a
particularly useful joint measurement which is fault tolerant \cite{FT}
in the sense that it operates even in the presence of errors occurring
during this measurement. An essential element for this measurement is
the introduction of a ``red light atom'' $R$ \cite{Scien} which reveals
the occurrence of errors. We will also show how to implement a universal
two--bit operation \cite{otherp} which also involves measurements that
indicate whether an error took place or not. In the former case, one has
to start the procedure again, whereas in the latter case, one knows one
has succeeded. Our schemes can be regarded as purification protocols
\cite{Purification} since with certain probability they are successful,
while sometimes the information is lost. We emphasize that in
applications in quantum communication the loss of information is not
central, whereas the knowledge that one has reliably transmitted the
quantum information is indispensable.

We start by discussing the physical details of our setup.
We consider two atoms, $1$ and $2$ inside a single cavity. The internal
structure of the atoms is displayed in Fig.\ 1; the qubit is stored in
the states $|0\rangle$ and $|1\rangle$, and there is an auxiliary state
$|r\rangle$. The states $|1\rangle$ and $|r\rangle$ are coupled by a
far--off--resonance Raman transition induced by an external laser field
and the cavity mode, whereas the state $|0\rangle$ is not coupled by
either the laser or the cavity field. The Hamiltonian describing the
interaction between the atoms and the cavity mode is given, in a
rotating frame at the cavity mode frequency, by
\begin{equation}
\label{Hamil}
H = \frac{g_1}{2} |1\rangle_{11}\!\langle r| a + 
\frac{g_2}{2} |1\rangle_{22}\!\langle r| a + h.c.
\end{equation}
where $a$ is the annihilation operator for the cavity mode, and
$g_{1,2}$ are the effective coupling constants of the Raman transition.
In the following, we will consider that a laser pulse of duration
$\Delta t_1=\pi/g_1$ is applied to atom $1$ and then another laser pulse
of duration $\Delta t_2=\pi/g_2$ is applied to atom $2$ \cite{adpass}.
Denoting by $|0\rangle_{\rm cav}$ and $|1\rangle_{\rm cav}$ the cavity
state of zero and one photons, respectively, this gives under ideal
conditions
\begin{mathletters}
\label{Onk}
\begin{eqnarray}
|0\rangle_1|0\rangle_2|0\rangle_{\rm cav}&\mapsto& 
|0\rangle_1|0\rangle_2|0\rangle_{\rm cav}\\
\label{Onkc}
|0\rangle_1|r\rangle_2|0\rangle_{\rm cav}&\mapsto& 
|0\rangle_1|r\rangle_2|0\rangle_{\rm cav}\\
|1\rangle_1|0\rangle_2|0\rangle_{\rm cav}&\mapsto& 
-i|r\rangle_1|0\rangle_2|1\rangle_{\rm cav}\\
|1\rangle_1|r\rangle_2|0\rangle_{\rm cav}&\mapsto& 
-|r\rangle_1|1\rangle_2|0\rangle_{\rm cav},
\end{eqnarray}
\end{mathletters}
where we have considered only the cases in which the first atom is in
$|0\rangle_1$ or $|1\rangle_1$ and the second atom is in $|0\rangle_2$
or $|r\rangle_2$, since this will be sufficient for our purposes. Note
that if the first atom is in the state $|0\rangle_1$ nothing will
change. However, if it is in $|1\rangle_1$, then it will be transferred
to $-i|r\rangle_1$. Then, if the second atom is in $|r\rangle_2$, then
it will be transferred to the state $-i|1\rangle_2$, whereas if it is in
$|0\rangle_2$, it will not change its state and a cavity photon will
remain in the cavity. In reality there will be errors. Since we are
considering a far--off resonance Raman transition, the most important
ones will be photon losses either at the mirrors or by leaking out of
the cavity. As in our previous work \cite{En97} we will also consider
systematic errors in the detuning, timing, laser pulses, phase shifts,
etc. It is straightfoward to account for these errors in Eq.\
(\ref{Onk}) by including the state of the environment and different
operators acting on it, as well as adding new terms in the last two
lines which describe the effect of photon loss (see below). On the other
hand, we will also need single--atom operations involving the three
atomic levels. As mentioned in the introduction, we will concentrate
here on errors occurring in processes involving two bits.

In the first part of this Letter, we will be interested in the following
situation: atom $2$ is initially in state $|0\rangle_2$, and is
transferred to state $|r\rangle_2$; then the process (\ref{Onk}) takes
place, followed by two single--atom operations, namely
$-|r\rangle_1\leftrightarrow |1\rangle_1$ and
$|r\rangle_2\leftrightarrow |0\rangle_2$ in the first and second atom,
respectively. Hence, ideally we have 
\begin{equation}
\label{ideal}
|0\rangle_1|0\rangle_2 \mapsto|0\rangle_1|0\rangle_2, \quad
|1\rangle_1|0\rangle_2 \mapsto|1\rangle_1|1\rangle_2. 
\end{equation}
In the presence of the errors mentioned above, 
\begin{mathletters}
\label{local}
\begin{eqnarray}
|0\rangle_1|0\rangle_2|1\rangle &\mapsto&
|0\rangle_1|0\rangle_2{\cal L}_0|1\rangle \\
|1\rangle_1|0\rangle_2|1\rangle &\mapsto&
|1\rangle_1|1\rangle_2{\cal L}_1|1\rangle +
|1\rangle_1|0\rangle_2{\cal L}_a|1\rangle 
\end{eqnarray}
\end{mathletters}
where $|E\rangle$ denotes the initial state of the environment
(including the cavity mode), and the operators ${\cal L}$ act on this
state. We used that one can optically pump the state $|r\rangle_1$ to
the state $|1\rangle_1$ after the whole procedure. Note that with this
notation this process is formally equivalent to the {\em photonic
channel} introduced in Ref.\ \cite{En97b}. 

In the following we will assume the environment operators ${\cal
L}_{0,1}$ fulfill the {\it stationary property} for two consecutive
operations 
\begin{equation}
\label{markov}
{\cal L}_1^{(2)}{\cal L}_0^{(1)}|E\rangle={\cal L}_0^{(2)}{\cal L}_1^{(1)}
|E\rangle,
\end{equation}
starting at times $t_{1,2}$, of duration $\Delta t_{1,2}$, respectively.
Here we have used the short hand notation ${\cal L}_i^{(j)}\equiv {\cal
L}_i(t_j,\Delta t_j)$, where $i=0,1$ and $j=1,2$. In Ref.\ \cite{En97},
the validity of (\ref{markov}) has been demonstrated for the present
model using the quantum trajectories approach. Here, as a simple
example, we illustrate this stationarity property in the context of
photon absorption: we consider a cavity mode coupled to a bath of
oscillators in the vacuum state $|E\rangle\equiv|{\bf 0}\rangle$ (i.e.,
at zero temperature). We assume a linear coupling Hamiltonian
\begin{equation}
H= \omega a^\dagger a + \sum_{k} \omega_k b_k^\dagger b_k   
+ \sum_k g_k (a^\dagger b_k + h.c.),
\end{equation}
where $b_k,b^\dagger_k$ are creation and annihilation operators for the
bath oscillators, and $\omega_k$ and $g_k$ the corresponding frequencies
and coupling constants. Denoting by $t$ the intial time, after a time $\Delta t$
we will have
\begin{eqnarray*}
|0\rangle_{\rm cav} |E\rangle &\rightarrow& |0\rangle_{\rm cav} |E\rangle,\\
&\equiv&  |0\rangle_{\rm cav} {\cal L}_0 (t,\Delta t)|E\rangle,\\
|1\rangle_{\rm cav} |E\rangle &\rightarrow& c(\Delta t)|1\rangle_{\rm cav}|E\rangle
+ |0\rangle_{\rm cav} \sum_k c_k(\Delta t) b_k^\dagger  |E\rangle,\\
&\equiv&  
|1\rangle_{\rm cav} {\cal L}_1(t,\Delta t)|E\rangle
+ |0\rangle_{\rm cav}  {\cal L}_a(t,\Delta t) |E\rangle.
\end{eqnarray*}
where $c$ and $c_k$ are c--numbers. Note that ${\cal L}_{0,1}$ only
depend on $\Delta t$ but not on the initial time $t$. Moreover, they
commute and therefore they satisfy (\ref{markov}). The stationary
property is related to the zero temperature of the reservoir, which for
optical frequencies is a good approximation even at room temperature. On
the other hand, one can verify that systematic errors also fulfill
(\ref{markov}) since the corresponding ${\cal L}_{0,1}$ will be
c--numberas only depending on $\Delta t$ but not on $t$.

Our goal is to use (\ref{local}) to perform ideal joint measurements and
entanglement operations as are required in quantum communication via a
photonic channel \cite{En97,En97b}. In this scheme, one has to perform a
local joint measurement on two atoms to check whether they are in the
state $|0\rangle|0\rangle$ or not. It must be implemented such that an
error occurring during this measurement will be detected by the
measurement itself. To be specific, let us consider two atoms in a state
$|\Psi\rangle = |\Psi_c\rangle |E_{c}\rangle + |0\rangle_1|0\rangle_2
|E_{a}\rangle$, where $|E_{c,a}\rangle$ denote unnormalized states of
the environment, and $|\Psi_c\rangle=\alpha|0\rangle_1 |1\rangle_2 +
\beta|1\rangle_1|0\rangle_2$ with $\alpha$ and $\beta$ arbitrary
coefficients. The goal is to make a filtering measurement of the state
$|0\rangle_1|0\rangle_2$, so that with certain probability the state of
the atoms is projected onto the $|\Psi_c\rangle$ which is the one we
want to keep intact. In order to perform the joint measurement we need
the red light atom, $R$, initially prepared in the state $|0\rangle_R$.
We use (\ref{local}) between atoms $1$ and $R$, and then between atoms
$2$ and $R$ [See Fig.\ 2(a)]. This gives the transformation
\begin{eqnarray}
|0\rangle_1|1\rangle_2|0\rangle_R&\mapsto& 
|0\rangle_1|1\rangle_2|1\rangle_R{\cal L}_1^{(2)}\!{\cal L}_0^{(1)}+\!
|0\rangle_1|1\rangle_2|0\rangle_R{\cal L}_a^{(2)}\!{\cal L}_0^{(1)}\nonumber\\
|1\rangle_1|0\rangle_2|0\rangle_R&\mapsto&
|1\rangle_1|0\rangle_2|1\rangle_R{\cal L}_0^{(2)}\!{\cal L}_1^{(1)}+\!
|1\rangle_1|0\rangle_2|0\rangle_R{\cal L}_0^{(2)}\!{\cal L}_a^{(1)}\nonumber\\
|0\rangle_1|0\rangle_2|0\rangle_R&\mapsto&
|0\rangle_1|0\rangle_2|0\rangle_R{\cal L}_0^{(2)}\!{\cal L}_0^{(1)}
\label{measurement}
\end{eqnarray}
where we have left out the state of the environment. Now a single--atom
measurement on atom $R$ in the state $|1\rangle_R$ or $|0\rangle_R$
reveals whether the joint state of atoms $1$ and $2$ was in the subspace
spanned by $|0\rangle_1|1\rangle_2$ and $|1\rangle_1|0\rangle_2$, or a
photon loss took place, respectively. In the first case, the state after
the measurement will become
\begin{eqnarray}
|\Psi\rangle &\mapsto& 
  (\alpha|0\rangle_1|1\rangle_2 {\cal L}_1^{(2)}{\cal L}_0^{(1)}
+ \beta|1\rangle_1|0\rangle_2 {\cal L}_0^{(2)}{\cal L}_1^{(1)})|E_{c}\rangle,\nonumber\\
 &=& |\Psi_c\rangle {\cal L}_1^{(2)}{\cal L}_0^{(1)} |E_{c}\rangle,
\end{eqnarray}
where we have used (\ref{markov}). We emphasize that the errors that may
occur during the joint measurement either factor out (operators ${\cal
L}_1$ and ${\cal L}_0$) or are projected out (terms containing ${\cal
L}_a$). 

Let us now show how this measurement can be used in the implementation
for quantum communication proposed in \cite{En97}. In that case one
needs the same three-level atoms, and the transmission between atom $1$
in the first node (cavity) and atom $2$ in the second node is performed
by using an appropriate laser pulse to transfer
$|1\rangle_1\mapsto|r\rangle_1$, producing one cavity photon. This
photon then travels to the second cavity, where it can induce the
inverse transition in a second atom, $|r\rangle_2\mapsto|1\rangle_2$ to
which the time inverse laser pulse is applied. Finally, we transfer
$|r\rangle_1 \mapsto |1\rangle_1$ in atom 1. Levels $|0\rangle_1$ and
$|0\rangle_2$ are not coupled by the laser field. Using the same
notation as before, this transmission can then be summarized as
(\ref{local}) but with local operators ${\cal L}$ replaced by the
corresponding transmission operators ${\cal T}$
\begin{mathletters}
\label{photon}
\begin{eqnarray}
|0\rangle_1|0\rangle_2 &\mapsto&
|0\rangle_1|0\rangle_2{\cal T}_0 \\
|1\rangle_1|0\rangle_2 &\mapsto&
|1\rangle_1|1\rangle_2{\cal T}_1 +
|1\rangle_1|0\rangle_2{\cal T}_a. 
\end{eqnarray}
\end{mathletters}
We expect that in any realistic situation $||{\cal T}_a|| > ||{\cal
L}_a||$. In \cite{En97} we showed how, using this channel, one can send
quantum information perfectly {\em provided local operations and
measurements are perfect}. Here we will show how to accomplish the same
goal using noisy local operations and the joint measurement described
above. We will concentrate on producing a distant EPR pair entangling
two atoms in different nodes [see Fig.\ 2(b)]. We consider one atom (1)
in the first cavity and two atoms (2 and $a$) in the second cavity.
Starting from state $|0\rangle_1+|1\rangle_1$ \cite{sqrt2}, we use the
channel (\ref{photon}) between atoms $1$ and $2$; then we interchange
$|0\rangle_1\leftrightarrow|1\rangle_1$ in atom $1$; then we use again
the channel (\ref{photon}) between atom $1$ and atom $a$; finally, we
reverse $|0\rangle_1\leftrightarrow|1\rangle_1$ in atom 1. Using this
procedure we obtain the map \cite{En97}
\begin{eqnarray}
\label{trans}
&&(|0\rangle_1+|1\rangle_1)|0\rangle_2|0\rangle_a \mapsto\\
&&(|0\rangle_1|0\rangle_2|1\rangle_a
+|1\rangle_1|1\rangle_2|0\rangle_a){\cal T}_1^{(2)}{\cal T}_0^{(1)}\nonumber\\
&&+|0\rangle_1|0\rangle_2|0\rangle_a{\cal T}_a^{(2)}{\cal T}_0^{(1)}
+|1\rangle_1|0\rangle_2|0\rangle_a{\cal T}_0^{(2)}{\cal T}_a^{(1)},\nonumber
\end{eqnarray}
where, as before, we have used the stationary property (see Ref.\ \cite{En97}), 
\begin{equation}
\label{markovt}
{\cal T}_1^{(2)}{\cal T}_0^{(1)}|1\rangle =
{\cal T}_0^{(2)}{\cal T}_1^{(1)}|1\rangle.
\end{equation}
The last two terms in (\ref{trans}) arise from photon loss errors, and
can be detected by performing a {\em joint measurement} on atoms 2 and
$a$, namely checking whether they are in the state
$|0\rangle_2|0\rangle_a$. In case they are not found in this state, a
single ion measurement on atom $a$ (in the basis $|0\rangle\pm
|1\rangle$) leaves atoms $1$ and $2$ in a maximally entangled state. The
joint measurement requires entanglement, and therefore is susceptible to
errors. However, we can use instead our implementation of this joint
measurement using the red light ion in cavity 2 [see Fig.\ 2(b)].
Repeating the transmission (\ref{trans}) and the subsequent measurement
(\ref{measurement}) until no photon loss was detected (the red light ion
is found in the state $|1\rangle_R$), yields, after having measured atom
$a$ in the basis $|0\rangle_a\pm|1\rangle_a$, the state
$|\psi\rangle_{12}=|0\rangle_1|0\rangle_2 \pm |1\rangle_1|1\rangle_2$.
With this EPR state one can already distribute a random secret key using
the Ekert protocol \cite{Ek91} for quantum cryptography \cite{fot}. 

For certain applications in quantum communication and quantum computing
a two--bit universal gate is required, since when combined with one--bit
operations this is sufficient for any unitary operation
\cite{applications}. This gate cannot be implemented using Eq.\
(\ref{local}) since there the state $|1\rangle|1\rangle$ is absent as
input state, whereas in the gate this state has to be present. We show
now how to perform the universal gate
\begin{mathletters}
\label{gate}
\begin{eqnarray}
|0\rangle_1|0\rangle_2\mapsto |0\rangle_1|0\rangle_2; &&
|1\rangle_1|0\rangle_2\mapsto -|1\rangle_1|0\rangle_2;\\
|0\rangle_1|1\rangle_2\mapsto |0\rangle_1|1\rangle_2;&&
|1\rangle_1|1\rangle_2\mapsto |1\rangle_1|1\rangle_2,
\end{eqnarray}
\end{mathletters}
with the present implementation in the presence of errors. The gate
consists of three steps: {\em (i)} A single atom operation on atom $2$
exchanges $|1\rangle_2\leftrightarrow|r\rangle_2$ while leaving the
state $|0\rangle_2$ unchanged; {\em (ii)} we perform a conditional
operation using the cavity mode such that the state
$|1\rangle_1|0\rangle_2\mapsto - |1\rangle_1|0\rangle_2$ by applying
(\ref{Onk}) twice; {\em (iii)} we apply the inverse of step (i). Note
that, according to the evolution given by (\ref{Hamil}), if the initial
state is $|1\rangle_1|0\rangle_2$ the cavity photon produced the first
time will be absorbed again by atom $1$ the second time, yielding a
minus sign, as desired. 

In reality there will be errors due to photon losses, phase shifts of the
states involved, and imperfect state transfer. 
After applying the gate one obtains, including these errors,
\begin{mathletters}
\label{gatei}
\begin{eqnarray}
|0\rangle_1|0\rangle_2&\mapsto& |0\rangle_1|0\rangle_2{\cal L}_{00}\\
|0\rangle_1|1\rangle_2&\mapsto& |0\rangle_1|1\rangle_2{\cal L}_{01}\\
|1\rangle_1|0\rangle_2&\mapsto& -|1\rangle_1|0\rangle_2{\cal L}_{10}
+\!|r\rangle_1|0\rangle_2{\cal L}_{r0}\\
|1\rangle_1|1\rangle_2&\mapsto& |1\rangle_1|1\rangle_2{\cal L}_{11}
+\!|r\rangle_1|1\rangle_2{\cal L}_{r1}+\!|r\rangle_1|r\rangle_2{\cal L}_{rr}.
\end{eqnarray}
\end{mathletters}
The ``photon loss'' errors ${\cal L}_{r0,r1,rr}$ can be detected by
measuring if the first atom is in state $|r\rangle_1$. In order to
perform the gate in the presence of all these errors we apply
(\ref{gatei}) four times but changing
$|0\rangle\leftrightarrow|1\rangle$ first in atom $1$, then in atom $2$
and again in atom $1$, after subsequent applications. Moreover, in the
last one we change the phase of the laser field acting on atom $2$ by
$\pi$ in the second part of step (ii) so that no extra minus sign is
added to the state $|1\rangle_1|0\rangle_2$ [therefore, this fourth
application performs just the (noisy) identity operation in order to
symmetrize the errors]. If no error is found during the whole procedure
(i.e. population in state $|r\rangle_1$) we obtain
\begin{mathletters}
\begin{eqnarray}
|0\rangle|0\rangle&\mapsto& |0\rangle|0\rangle
{\cal L}_{01}^{(4)}{\cal L}_{11}^{(3)}{\cal L}_{10}^{(2)}{\cal L}_{00}^{(1)}\\
|0\rangle|1\rangle&\mapsto& |0\rangle|1\rangle
{\cal L}_{00}^{(4)}{\cal L}_{10}^{(3)}{\cal L}_{11}^{(2)}{\cal L}_{01}^{(1)}\\
|1\rangle|0\rangle&\mapsto& -|1\rangle|0\rangle
{\cal L}_{11}^{(4)}{\cal L}_{01}^{(3)}{\cal L}_{00}^{(2)}{\cal L}_{10}^{(1)}\\
|1\rangle|1\rangle&\mapsto& |1\rangle|1\rangle
{\cal L}_{10}^{(4)}{\cal L}_{00}^{(3)}{\cal L}_{01}^{(2)}{\cal L}_{11}^{(1)}.
\end{eqnarray}
\end{mathletters}
Using the same arguments as in (\ref{markov}), one can check that all
these operators are identical. Thus, once no error was found the gate
worked perfectly.

So far, we used the stationary properties (\ref{markovt}) and
(\ref{markov}) for transmission and local operations. It is important to
realize that, even if the former one (\ref{markovt}) does not hold, one
can still establish a perfect EPR pair, since we have shown here how to
purify all local operations (including the gate) needed for the
procedure developed in \cite{En97b}. On the other hand, if also
(\ref{markov}) would not hold, one can establish an entangled state
whose degree of entanglement is limited by the degree to which
(\ref{markov}) is satisfied.

In summary, we have shown how perform joint measurements in the presence
of errors in a cavity QED implementation. The scheme works even if
errors occur during the measurement itself. We have shown how to apply
this proposal in quantum communication to achieve perfect transmission
over a nosiy channel including local errors. Using the same
implementation, we have also presented a universal two--bit gate that
operates perfectly in the presence of errors.

This work was supported in part by the TMR network
ERB--FMRX--CT96--0087, and by the Austrian Science Foundation.

% -------------------------------------------------------------------------

% -------------------------------------------------------------

\begin{figure}[htbp]
\begin{center}
   \leavevmode
      \epsfxsize=8cm  \epsfbox{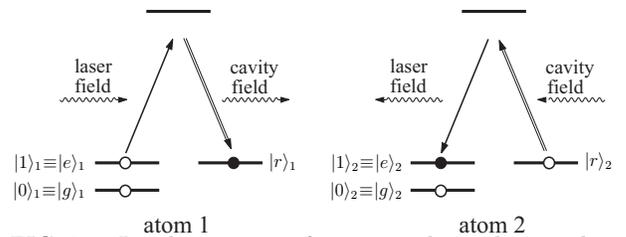}  
\caption{ Level
structure of atoms and couplings induced by laser and cavity fields.
}
\end{center}
\end{figure}

\begin{figure}[htbp]
\begin{center}
   \leavevmode
      \epsfxsize=8cm  \epsfbox{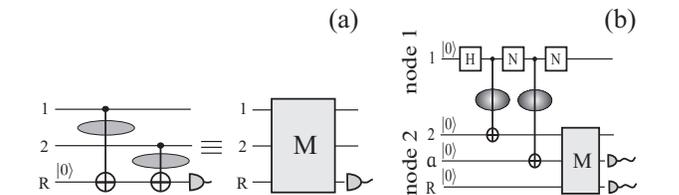}  
\caption{Diagrammatic representation of: (a) Joint measurement;
(b) Establishing an EPR pair. H and N denote the Hadamard 
and Not transformations, respectively.
}
\end{center}
\end{figure}


\begin{references}

\bibitem{Bell}
J.S. Bell,
Physics {\bf 1}, 195 (1965).

\bibitem{applications} C. H. Bennett, Phys.~Today, volume 24
	(October 1995); D. P. DiVincenzo, Science {\bf 270}, 255 (1995).

\bibitem{Wi95}
C. Monroe {\em et al.}, Phys. Rev. Lett. {\bf 75},
4714 (1995); for a theoretical proposal see
J.I. Cirac, and P.~Zoller,
Phys. Rev. Lett., {\bf 74}, 4091 (1995).

\bibitem{Kimble}
Q. Turchette {\em et al.}, Phys. Rev. Lett. {\bf 75}, 4710 (1995).

\bibitem{Ha97} Entanglement of two atoms has been very recently
achieved for the first time, 
E. Hagley {\em et al.}, Phys. Rev. Lett. {\bf 79}, 1
(1997). 

\bibitem{QCryptoexp}
P.D. Townsend, Electron. Lett. {\bf 30}, 809 (1994) and {\em ibid.}
{\bf 33}, 188 (1997);
J.D. Franson and H. Ilves, Appl. Optics {\bf 33}, 2949 (1994);
A. Muller, H. Zbinden and N. Gisin,
Europhys. Lett. {\bf 33}, 335 (1995);

\bibitem{QN} J.I. Cirac {\em et al.}, Phys. Rev. Lett. {\bf 78}, 3221
(1997); see also T. Pellizari, quant-ph/9707001.

\bibitem{En97}S. J. van Enk, J. I. Cirac and P. Zoller,
Phys. Rev. Lett. {\bf 78}, 4293 (1997).

\bibitem{qerrorc}
P.W. Shor, Phys. Rev. A {\bf 52}, 2493 (1995);
A.M. Steane, Phys. Rev. Lett. {\bf 77}, 793 (1996);
E. Knill and R. Laflamme, Phys. Rev. A  {\bf 55}, 900 (1997).

\bibitem{Scien}
J.I. Cirac, T. Pellizzari, and P. Zoller, Science {\bf 273}, 1207
(1996).

\bibitem{Purification}
C.H. Bennett {\it et al}, Phys. Rev. Lett. {\bf 76}, 722 (1996);
A. Ekert and C. Macchiavello, {\it ibid}, {\bf 77}, 2585 (1996).

\bibitem{Schumacher}
B. Schumacher, Phys. Rev. A {\bf 45}, 2614 (1996);
C.H. Bennett, D.P. DiVincenzo and J.A. Smolin,
Phys. Rev. Lett. {\bf 78}, 3217 (1997).

\bibitem{FT}
P. Shor, quant-ph/9605011;
J. Preskill, quant-ph/9705031;
E. Knill, R. Laflamme, and W.H. Zurek, quant-ph/9702058.


\bibitem{otherp}Other proposals for implementing a 
2--bit gate include
P. Domokos {et al.}, Phys. Rev. A {\bf 52}, 3554 (1995);
T. Pellizari {et al.}, Phys. Rev. Lett. {\bf 75}, 3788 (1995).

\bibitem{adpass}
One can also use the adiabatic passage technique
and apply simultaneously both laser pulses giving the same result 
(see \protect\cite{otherp}).

\bibitem{En97b}S. J. van Enk, J. I. Cirac and P. Zoller, unpublished.

\bibitem{sqrt2} 
For simplicity we omit trivial normalization factors.

\bibitem{Ek91}A. Ekert,
Phys. Rev. Lett. {\bf 67}, 661, (1991).

%\bibitem{QCrypto}
%C. H. Bennett, Phys. Today, Vol. 24 (October 1995). 

\bibitem{fot}
Using similar ideas, one can realize the local entanglement operations
needed to send arbitrary states in the protocol presented in Ref.\ \cite{En97}.


\end{references}
\end{document}